\documentclass[a4paper,twocolumn,11pt]{quantumarticle}
\pdfoutput=1
\usepackage[utf8]{inputenc}
\usepackage[english]{babel}
\usepackage[T1]{fontenc}
\usepackage{amsmath}
\usepackage{hyperref}

\usepackage{tikz}
\usepackage{lipsum}

\usepackage{multirow}

\begin{document}

\title{Computing the gradients with respect to all parameters of a quantum
neural network using a single circuit}

\author{Guang Ping He}
\affiliation{School of Physics, Sun Yat-sen University, Guangzhou 510275, China}
\email{hegp@mail.sysu.edu.cn}
\orcid{0000-0002-3541-409X}
\maketitle

\begin{abstract}
Finding gradients is a crucial step in training machine learning models. For quantum neural networks, computing gradients using the parameter-shift rule requires calculating the cost function twice for each adjustable parameter in the network. When the total number of parameters is large, the quantum circuit must be repeatedly adjusted and executed, leading to significant computational overhead. Here we propose an approach to compute all gradients using a single circuit only, significantly reducing both the circuit depth and the number of classical registers required. We experimentally validate our approach on both quantum simulators and IBM's real quantum hardware, demonstrating that our method significantly reduces circuit compilation time compared to the conventional approach, resulting in a substantial speedup in total runtime.
\end{abstract}

%


\section{Introduction}

Artificial intelligence technology is making incredible progress nowadays.
One of the main reason is that classical artificial neural networks are made
very feasible with the advance of computer science in recent years. On the
other hand, though it is widely believed that quantum neural networks may
breathe new life into the researches \cite{ml179}, they are still not as
practical as their classical counterparts due to the scale and performance
of currently available quantum computers. Especially, highly effective
methods have been developed for computing the gradients of the cost function
with respect to the adjustable parameters of classical neural networks,
e.g., the backpropagation algorithm \cite{back,ml98}. A distinct feature is
that the cost function needs to be calculated only once, and all the
gradients will be readily deduced, so that classical neural networks can be
trained fast and efficiently. But for quantum neural networks, computing the
gradients is considerably less convenient \cite{ml211}. Effective algorithms
were found for certain network architectures only \cite{ml209}. For other
general architectures, the parameter-shift rule \cite{ml124,ml138} seems to
be the best algorithm found for this task so far. But it needs to calculate
the cost function twice for computing the gradient with respect to a single
parameter of the network (which will be further elaborated below). The
values of some parameters for each calculation need adjustment too.
Consequently, the quantum circuit for calculating the cost function has to
be modified many times in order to obtain the gradients with respect to all
parameters in each single round of training of the quantum network.

In this paper, we will propose an approach in which the gradients with
respect to all parameters of a quantum neural network can be computed
simultaneously using a single circuit only. The main purpose is to break the
limit on the scale of computation on real quantum hardware. As it is
well-known, the number of adjustable parameters of a powerful neural network
is generally very high. If we have to run two quantum circuits to compute
the gradient with respect to a single parameter, then the total number of
unique executed circuits could easily exceed the capacity of any real
quantum computer in existence. Of course, the execution of these circuits
can be broken down into many \textquotedblleft jobs\textquotedblright . But
it is obviously inconvenient when we are using quantum cloud computation
platforms. Especially, if the jobs are submitted separately, then there
could be a long waiting time between the execution of the jobs. For example,
for free users on the IBM Quantum Experience online platform, this waiting
time could vary from $10$ minutes to $4$ hours. For paying users, as also
pointed out in Ref. \cite{ml138}, \textquotedblleft quantum hardware devices
are typically billed and queued per unique circuit\textquotedblright .
According to Amazon Braket, the quantum computation platforms Rigetti, IonQ,
OQC, and Quera all charge \$0.3 USD per unique circuit in addition to other
fees \cite{amazon}. A conventional method for reducing the number of unique
circuits is to \textquotedblleft stack\textquotedblright\ many circuits into
a single job. In the Qiskit software development kit for running IBM quantum
computers and simulators with Python programs, this can be done using the
\textquotedblleft append\textquotedblright\ function \cite{example}. But
there is also a restriction on its usage on real quantum computers. For
example, on the \textquotedblleft ibmq\_quito\textquotedblright\ backend,
the number of circuits allowed to be appended is limited to $100$. To break
this limit, the Qiskit's \textquotedblleft compose\textquotedblright\
function can serve as an alternative. According to our experience, it runs
much slower than the \textquotedblleft append\textquotedblright\ method
does. On platforms that charge by the runtime, slower programs mean higher
cost. As of now, the rate in IBM Cloud's Pay-As-You-Go Plan is \$96 USD
per minute \cite{IBM}. Nevertheless, it can indeed stack more circuits into
one single job so that the gradients with respect to more parameters can be
computed together.

In the following, we will use the direct stacking method based on the
\textquotedblleft compose\textquotedblright\ function (referred to as the
\textquotedblleft conventional approach\textquotedblright ) as a baseline to
compare with the performance of our proposed method that uses a single
circuit only (referred to as the \textquotedblleft improved
approach\textquotedblright ). In the next section, we will briefly introduce
the typical structure of quantum neural networks. The parameter-shift
rule will be reviewed in Section 3. Subsequently, our improved approach will
be presented in Section 4, with its theoretical advantages --- namely,
reduced circuit depth and lower classical memory requirements --- elaborated
in Section 5. However, this approach also carries some theoretical
disadvantages: the circuit needs two additional qubits, and requires more shots to run compared to the
conventional approach, as explained in Section 6. To demonstrate that the
advantages outweigh these disadvantages, we conduct experiments, and the
results are reported in Section 7. These results show a significant speedup
in total runtime, particularly as the number of input data points and the
scale of the quantum circuit increase. Finally, in Section 8, we discuss the
generalization of our approach, and suggest some improvements on real quantum hardware which can make our approach even faster.



\section{The variational quantum circuit}


\begin{figure}[tbp]
\centering
\includegraphics[scale=0.8]{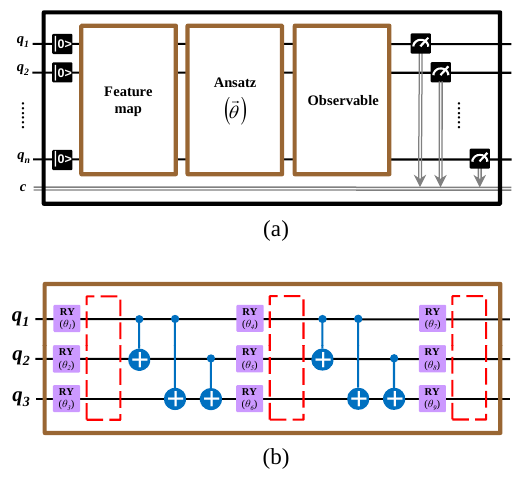}
\caption{(a) The general structure of a variational quantum circuit (VQC).
All qubits $q_i$ ($i=1,...,n$) are initialized in the state $\left\vert
0\right\rangle $, then pass through three blocks: the feature map, the
ansatz and the observable, which apply certain unitary transformations.
Finally they are measured in the computational basis, and the result is sent
to the classical register $c$. All the adjustable parameters $\vec{\protect%
\theta}=(\protect\theta _{1},...\protect\theta _{n})$ are contained in the
ansatz. (b) An example of the \textit{RealAmplitudes} ansatz on three
qubits, with two repetitions and full entanglement. $RY(\protect\theta _{i})$
($i=1,...,9$) are single-qubit RY rotation gates, with the adjustable
parameter $\protect\theta _{i}$\ denoting the rotation angle about the $y$%
-axis. The blue icons are CX (controlled-NOT) gates. The red dashed boxes
denote where the additional circuit of our improved approach shown in Fig.3
will be added.}
\label{fig:epsart}
\end{figure}


There are various types of quantum neural networks. Among them, variational
quantum circuits (VQCs) \cite{ml128,ml383,ml54,ml118,ml52,ml53,ml32,ml208}
have the advantage that it can be implemented on current or near-term noisy
intermediate-scale quantum computers, as demonstrated experimentally \cite%
{ml368,ml157,ml367,ml366}. Applications within this framework also include
variational quantum simulators \cite{ml382,ml381}, quantum approximate
optimization algorithm \cite{ml30,ml380,ml387}, et al. Research domains
extend across high-energy physics \cite{161-of-ml224,ml379}, cybersecurity
\cite{159-of-ml224,ml389}, and finance \cite{160-of-ml224,ml386}. In this
paper, we will use VQCs as an example to illustrate how our improved
approach can be applied to compute the gradients with respect to their
parameters.

As shown in Fig.1(a), the general structure of a VQC can be divided into
three blocks: the feature map, the ansatz, and the measurement of the
observable, which gives the value of the cost function. Note that the goal
of this paper is to provide a method for computing the gradients, instead of
finding a VQC for a specific application. To this end, the details of the
feature map and the measurement are less relevant, and our approach does not
need to modify them. Thus, we will leave their description in the
Experiments section. The ansatz is the most important block, because it
contains all the adjustable parameters with respect to whom the gradients
need to be computed. Our example uses the \textit{RealAmplitudes} ansatz
\cite{ansatz}, which is the default ansatz in Qiskit's VQC implementation.
As pointed out in Ref. \cite{ml32}, this ansatz has also been used in a VQC
with proven advantages over traditional feedforward neural networks in terms
of both capacity and trainability \cite{ml53}. The typical structure of the
\textit{RealAmplitudes} ansatz is illustrated in Fig.1(b).


\section{The parameter-shift rule}

Finding the gradients is a must for optimizing neural networks for certain
applications. In our example, the original parameter-shift rule proposed in
Ref. \cite{ml124} will be used, where the exact gradient of the cost
function $f=f(\vec{\theta})$ with respect to the parameters $\vec{\theta}%
=(\theta _{1},...\theta _{n})$ of the quantum gates is obtained as%
\begin{equation}
\frac{\partial f}{\partial \theta _{i}}=r\left( f\left( \vec{\theta}+s\vec{e}%
_{i}\right) -f\left( \vec{\theta}-s\vec{e}_{i}\right) \right)  \label{rule}
\end{equation}%
for $i=1,...,n$, where $r=1/2$ and $s=\pi /2$ for all the rotation and phase
gates available in the Qiskit library.

This equation shows that computing the gradient with respect to a single
parameter $\theta _{i}$\ needs to calculate the cost function $f$ twice,
each time with a different set of parameters, i.e., $\vec{\theta}+s\vec{e}%
_{i}=(\theta _{1},...,\theta _{i}+s,...\theta _{n})$\ and $\vec{\theta}-s%
\vec{e}_{i}=(\theta _{1},...,\theta _{i}-s,...\theta _{n})$. Therefore, when
the total number of parameters of a VQC is $n$, computing all the gradients
will have to run $2n$ circuits with different values of the parameters.
To stack these circuits into a single job, the aforementioned conventional
approach is simply to connect them one after another using Qiskit's
\textquotedblleft compose\textquotedblright\ function, as illustrated in
Fig.2. Each circuit is independently executed and measured before the next
one is re-initialized and parameterized with new values, and the measurement
results are stored in different classical registers.


\begin{figure}[bt]
\centering
\includegraphics[scale=0.6]{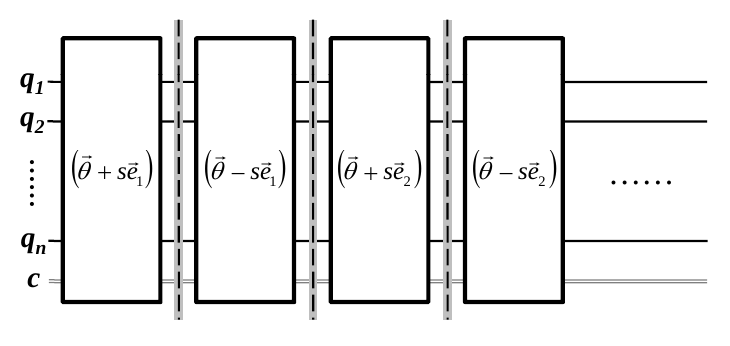}
\caption{Diagram of the conventional approach, in which the circuits of many
VQCs are stacked in serial using Qiskit's \textquotedblleft
compose\textquotedblright\ function. Each black box contains all the
components of the complete VQC shown in Fig.1(a), i.e., the feature map, the
ansatz, and the observable along with the measurement. But the adjustable
parameters $\vec{\protect\theta}=(\protect\theta _{1},...\protect\theta %
_{n}) $ for each ansatz take different values.}
\label{fig:epsart}
\end{figure}


\section{Our improved approach}



\begin{figure*}[thbp]
\centering
\includegraphics[scale=0.8]{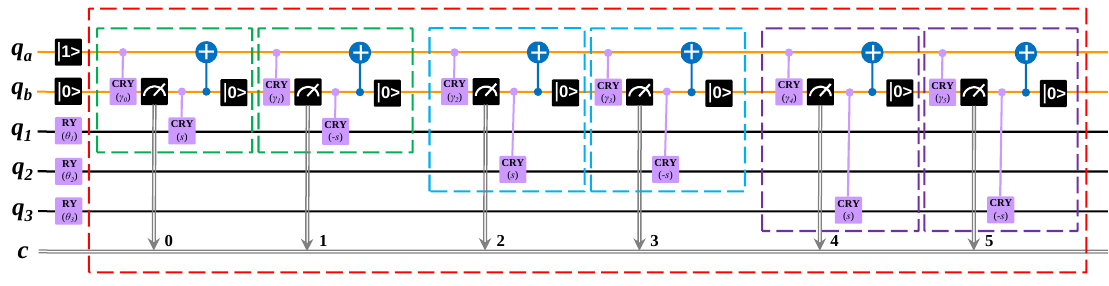}
\caption{The quantum circuit for our improve approach. The section inside
the red dashed box can be put into any one of the three red dashed boxes in
Fig.1(b). Here $q_{a}$ and $q_{b}$\ are the additional control qubits. They
need to be initialized in the states $\left\vert 1\right\rangle _{q_{a}}$
and $\left\vert 0\right\rangle _{q_{b}}$ only once at the very beginning of
Fig.1(b). The two green/blue/purple dashed boxes will create the shift $\pm
s $ to the gates $RY(\protect\theta _{1})/RY(\protect\theta _{2})/RY(\protect%
\theta _{3})$, respectively, with equal probabilities as controlled by $%
q_{a} $ and $q_{b}$. Each of these dashed boxes starts with a controlled-RY
gate $CRY(\protect\gamma _{j})$, where $\protect\gamma _{j}$ is defined in
Eq. (\protect\ref{gamma j}). Then a measurement is made on $q_{b}$, with the
result recorded in the classical registers. Another controlled-RY gate $%
CRY(\pm s)$ is placed between $q_{b}$ and $q_{i}$ ($i=1,2,3$), followed by a
CX gate between $q_{b}$ and $q_{a}$. Finally, $q_{b} $\ is reset to the
state $\left\vert 0\right\rangle _{q_{b}}$.}
\label{fig:epsart}
\end{figure*}


Now it will be shown how to compute all the $2n$ shifted\ cost functions as
well as the original unshifted\ cost function $f(\vec{\theta})$\ using a
single circuit. The main idea is to introduce additional gates for realizing
the shifts to the existing parameterized gates of the ansatz. Each of these
additional gates is activated with a certain probability, so that each shift
stands a chance to take effect. When none of these gates activate, the
circuit acts exactly like the original ansatz so that the unshifted\ cost
function will be computed. The key part is to ensure that these gates will
be activated only one at a time at the most. That is, in each run (i.e.,
\textquotedblleft shot\textquotedblright ) of the circuit, there should not
be two (or more) additional gates activate simultaneously. Otherwise, the
circuit will output the cost function with the form like $f\left( \vec{\theta%
}+s\vec{e}_{i}+s\vec{e}_{j}\right) $ ($i\neq j$) which could be useful for
calculating higher-order derivatives but is useless for computing the
gradient via Eq. (\ref{rule}), while significantly lowers the efficiency of
the circuit.

For this purpose, we add two additional control qubits $q_{a}$ and $q_{b}$\
to the circuit, regardless the number of qubits and parameters in the
original ansatz. Then for each existing single-parameter rotation gate $%
RY(\theta _{i})$ ($i=1,...,n$), we added two blocks acting on the qubit it
rotates and the above two additional control qubits, to turn the parameter
from $\theta _{i}$\ to $\theta _{i}+s$\ and $\theta _{i}-s$, respectively.
Each block contains $2$ controlled-RY gates, $1$ CX (two-qubit
controlled-NOT) gate, $1$ measurement and $1$ reset operation. Fig.3
showcases an example of the resultant circuit. For simplicity, only three RY
gates in the \textit{RealAmplitudes} ansatz of the original VQC
(corresponding to the three RY gates before any of the red dashed boxes in
Fig.1(b)) are studied, and shown with the two additional blocks added to
each of them. We also uploaded the diagram of the complete quantum circuit
for a VQC with $10$ qubits (which will be used in the Experiments section
for the classification task of the MNIST dataset) to Github (available at:
https:// github.com/gphehub/grad2210/blob/main/ Fig.7\_Complete circuit for
MNIST classification.emf).

Let us track the circuit in Fig.3 from left to right to see how it works.
The role of the first additional control qubit $q_{a}$ is to record whether
any of the additional blocks has been activated. It can also be considered
as a switch. If it is in the state $\left\vert \psi \right\rangle
_{q_{a}}=\left\vert 1\right\rangle _{q_{a}}$, then it means that no block
has ever been activated yet. Or if it is flipped to the state $\left\vert
\psi \right\rangle _{q_{a}}=\left\vert 0\right\rangle _{q_{a}}$, then it
implies that one of the blocks was already activated. The second additional
control qubit $q_{b}$ acts like a quantum dice, which controls the
probability for each block to be activated.
At the very beginning of the whole VQC, 
$q_{a}$ and $q_{b}$ are initialized in the states $\left\vert 1\right\rangle
_{q_{a}}$ and $\left\vert 0\right\rangle _{q_{b}}$, respectively.

Next, in the first green block right behind the first RY gate $RY(\theta
_{1})$, the first gate $CRY(\gamma _{0})$ is a controlled-RY gate with $%
q_{a} $\ ($q_{b}$) serving as the control (target) qubit. It introduces a
rotation angle%
\begin{equation}
\gamma _{0}=2\arcsin \sqrt{\frac{1}{N-0}}
\end{equation}%
about the $y$-axis to $q_{b}$\ only when $\left\vert \psi \right\rangle
_{q_{a}}=\left\vert 1\right\rangle _{q_{a}}$. Here $N=2n+1$. Then the state
of $q_{b}$\ becomes%
\begin{eqnarray}
\left\vert \psi \right\rangle _{q_{b}} &=&\cos \frac{\gamma _{0}}{2}%
\left\vert 0\right\rangle _{q_{b}}+\sin \frac{\gamma _{0}}{2}\left\vert
1\right\rangle _{q_{b}}  \nonumber \\
&=&\sqrt{1-\frac{1}{N-0}}\left\vert 0\right\rangle _{q_{b}}+\sqrt{\frac{1}{%
N-0}}\left\vert 1\right\rangle _{q_{b}}.  \label{qb}
\end{eqnarray}%
A measurement in the computational basis $\{\left\vert 0\right\rangle
,\left\vert 1\right\rangle \}$ is then made on $q_{b}$, and the result is
sent to a classical register. With probability $1/N$\ the measurement result
will be $\left\vert 1\right\rangle _{q_{b}}$. In this case, the
controlled-RY gate $CRY(s)$ next to the measurement will be activated,
introducing a rotation angle $s=\pi /2$ to $q_{1}$\ (i.e., the qubit on
which the gate $RY(\theta _{1})$\ of the original ansatz is applied). Since
the RY gates has the property%
\begin{equation}
RY(s)RY(\theta _{1})=RY(\theta _{1}+s),
\end{equation}%
we can see that a shift $s$ is successfully introduced to the parameter $%
\theta _{1}$. Meanwhile, with the next CX gate where $q_{b}$\ ($q_{a}$)
serves as the control (target) qubit, the state of $q_{a}$\ is turned from $%
\left\vert 1\right\rangle _{q_{a}}$\ into $\left\vert 0\right\rangle
_{q_{a}} $, indicating that the current block was activated. Also, at the
end of this block, the state of $q_{b}$ is reset to $\left\vert
0\right\rangle _{q_{b}}$ regardless the measurement result. Consequently,
all the rest blocks that follows will be bypassed. After executing the
entire circuit, the final result will give the cost function $f\left( \vec{%
\theta}+s\vec{e}_{1}\right) $.

On the other hand, with probability $1-(1/N)$ the measurement result on $%
q_{b}$ in this block will be $\left\vert 0\right\rangle _{q_{b}}$, as
indicated by Eq. (\ref{qb}). Then the two aforementioned controlled gates in
the same block following this measurement will not take effect at all, so
that the states of $q_{a}$\ and $q_{1}$\ remain unchanged, allowing the
second green block to be activated.

Similarly, in the second green block, since the state of $q_{b}$ was reset
to $\left\vert 0\right\rangle _{q_{b}}$, the first gate $CRY(\gamma _{1})$
introduces a rotation angle%
\begin{equation}
\gamma _{1}=2\arcsin \sqrt{\frac{1}{N-1}}
\end{equation}%
to $q_{b}$\ when $\left\vert \psi \right\rangle _{q_{a}}=\left\vert
1\right\rangle _{q_{a}}$, turning the state of $q_{b}$\ into%
\begin{eqnarray}
\left\vert \psi \right\rangle _{q_{b}} &=&\cos \frac{\gamma _{1}}{2}%
\left\vert 0\right\rangle _{q_{b}}+\sin \frac{\gamma _{1}}{2}\left\vert
1\right\rangle _{q_{b}}  \nonumber \\
&=&\sqrt{1-\frac{1}{N-1}}\left\vert 0\right\rangle _{q_{b}}+\sqrt{\frac{1}{%
N-1}}\left\vert 1\right\rangle _{q_{b}}.
\end{eqnarray}%
Then $q_{b}$\ is also measured in the computational basis, and the result is
stored in another classical register. With probability $1/(N-1)$ the result
will be $\left\vert 1\right\rangle _{q_{b}}$. Note that this case occurs
only when the first block was not activated, i.e., the measurement result of
$q_{b}$\ in the first block was $\left\vert 0\right\rangle _{q_{b}}$, which
occurred with probability $1-(1/N)$. Thus, the total probability for finding
$q_{b}$\ in $\left\vert 1\right\rangle _{q_{b}}$ at this stage (i.e., the
second block is activated) is%
\begin{equation}
prob_{2}=\left( 1-\frac{1}{N}\right) \frac{1}{N-1}=\frac{1}{N},
\end{equation}%
which equals to the probability for activating the first block. When this
result indeed occurs, the next gate $CRY(-s)$ will introduce the rotation
angle $-s=-\pi /2$\ to $q_{1}$, and the CX gate next to it will turn the
state of $q_{a}$\ into $\left\vert 0\right\rangle _{q_{a}}$, making the rest
blocks bypassed. Then the final result of the entire circuit will be the
cost function $f\left( \vec{\theta}-s\vec{e}_{1}\right) $.

For the same reasons, by setting the rotation angle of the first
controlled-RY gate $CRY(\gamma _{j})$\ in the $(j+1)$th additional block as%
\begin{equation}
\gamma _{j}=2\arcsin \sqrt{\frac{1}{N-j}}  \label{gamma j}
\end{equation}%
($j=0,...,N-2$), it can be proven that every block stands the same
probability $1/N$\ to be activated. The CX gates on $q_{b}$\ and $q_{a}$\
also ensure that only one block at the most will be activated. As a result,
executing the whole circuit for one shot will give the output corresponding
to the shifted cost function $f\left( \vec{\theta}+s\vec{e}_{j}\right) $ or $%
f\left( \vec{\theta}-s\vec{e}_{j}\right) $\ for a single $j$ only. When none
of the block was activated (which also occurs with probability $1/N$), the
circuit outputs the unshifted cost function $f\left( \vec{\theta}\right) $.
By reading the classical registers $c$ that store the measurement results of
$q_{b}$ in every block to see which block was activated, we can learn which
cost function was calculated. Running the circuit for many shots will then
provide enough shots for each of these cost functions, so that all gradients
in Eq. (\ref{rule}) can be obtained with the desired precision.

After finding the gradients, the standard training routine can be applied
for optimizing the corresponding VQC. That is, the value of each adjustable
parameter $\theta _{i}$ can be updated as%
\begin{equation}
\theta _{i}\rightarrow \theta _{i}^{\prime }=\theta _{i}-\eta \frac{\partial
f}{\partial \theta _{i}},
\end{equation}%
where $\eta $ is the learning rate chosen by the user. Then the VQC should
be reconstructed by using $\vec{\theta}^{\prime }=(\theta _{1}^{\prime
},...\theta _{n}^{\prime })$ as the new adjustable parameters of the ansatz.
This completes one epoch of training. Computing the gradients using the new
VQC and repeating this updating procedure for many epochs will eventually
minimize the unshifted cost function, which means that the VQC is optimized
for the given task.

\section{Theoretical advantages}


\subsection{Circuit depth}


Comparing with the conventional approach, a significant advantage of our
improved approach is that the depth of the entire circuit is reduced, as
estimated below.

For each single input data, let $\Lambda $\ denote the depth of the circuit
for calculating the cost function once using the conventional approach, and $%
n$ denote the number of the adjustable parameters. Define%
\begin{equation}
\lambda =\frac{\Lambda }{n}.  \label{lamda}
\end{equation}%
In many ansatz, $\lambda $ remains non-trivial even for high $n$ (which will
be proven in 
Appendix A by using the \textit{RealAmplitudes} ansatz as an example).
Recall that computing the gradient with respect to a single parameter using
the parameter-shift rule needs to calculate the cost function twice. Thus,
the total depth of the stacked circuit for computing all the $n$ gradients
and the unshifted cost function using the conventional approach is%
\begin{equation}
D_{conv}=\Lambda (2n+1)=\lambda n(2n+1)\propto O(n^{2}).
\end{equation}

On the other hand, when using our improved approach, from Fig.3\ we can see
that a total of $10$ operations were added to each existing parameterized RY
gate of the original ansatz.
Therefore, to compute all the $n$ gradients and the unshifted cost function,
the total depth of our circuit is simply%
\begin{equation}
D_{impr}=\Lambda +10n=\lambda n+10n\propto O(n).  \label{improved depth}
\end{equation}%
This result shows that our improved approach has the advantage that it takes
a lower circuit depth when the number of parameters $n$ is high, and the
improvement will be more significant with the increase of $n$.

\subsection{Number of classical bits in the stacked circuits}



In the conventional approach, for each single input data, running the
circuit for calculating the cost function once (either shifted or unshifted)
takes $Q$ classical bits to store the final measurement result of the $Q$
qubits. When all the $2n+1$ circuits for computing all the $n$ gradients and
the unshifted cost function are stacked together so that it can be submitted
as a single job on the real hardware, the total number of required classical
bits is{}%
\begin{equation}
N_{conv}=Q(2n+1).
\end{equation}

In our improved approach, the $Q$ qubits of the original ansatz are measured
only once at the very end of the circuit. The additional qubit $q_{b}$ is
measured twice in the two additional blocks added to each of the $n$
parameterized gates of the original ansatz, which takes $2n$ classical bits
to store the results. At the end of the circuit, both the additional qubits $%
q_{a}$\ and $q_{b}$ do not need to be measured. But for simplicity, we added
these two measurements in our code too (not shown in Fig.3), which may also
serve as an additional monitoring of the running of the circuit.
Consequently, the total number of classical bits required in our approach
is{}%
\begin{equation}
N_{impr}=Q+2n+2,
\end{equation}%
which is smaller than that of the conventional approach since generally $%
Q\geq 2$. This is another advantage of our approach.

\section{Theoretical disadvantages}


Our improved approach requires two more qubits than the conventional
approach. It is considerable on free quantum computation platforms, which
generally offers a total of $5\symbol{126}6$ qubits only. But on near-term
quantum devices that could really play a role in practical applications, the
number of qubits has to be much higher, so that taking two more qubits
should not make much difference.

What really matters is that our approach takes more shots than the
conventional approach does 
in order to compute the gradients up to approximately the same precision.
This is because in the conventional approach, when the stacked circuit is
executed for $s$ shots, each of the shifted and unshifted cost functions is
calculated for exactly $s $ shots too. But in our approach, when executing
the circuit once, only one of the cost functions is calculated, depending on
which of the additional blocks was activated by chance. To ensure that each
cost function will be calculated for approximately $s$ shots, our circuit
needs to be executed for about $s^{\prime }=s(2n+1)$ shots in total. This
surely increases the runtime of the program. But recall that our circuit has
a lower depth, so that it takes less time to compile. Therefore, whether our
approach can result in a speedup on the total runtime depends on the
competition between these two factors, and will be studied experimentally
below.

Also, in our approach whether an additional block will be activated is
controlled by the measurement result of the qubit $q_{b}$, where quantum
uncertainty takes effect. Consequently, executing the circuit for $s^{\prime
}=s(2n+1)$ shots does not mean that each cost function will be calculated
for $s$ shots exactly. Some statistical fluctuations are inevitable.
Nevertheless, the relationship between the number of shots and the precision
of the results is also statistical and subjected to quantum uncertainty. A
fluctuation of the precision cannot be avoided even if all individual shot
numbers are strictly aligned to the same value. Thus, we do not see the need
to take extra efforts trying to level off the individual shot numbers of all
the cost functions so that they equal exactly to each other. (In fact, this can simply be accomplished by discarding the data of some of the shots. But like we said, it seems unnecessary so that our experiments do not include this treatment.)


\section{Experiments}

To compare the performance of our improved approach and the conventional
one, we tested them experimentally on both real quantum hardware and
simulator.

\subsection{The input datasets}


Two classical input datasets are used in our experiments. One of them is an $%
8 $-dimensional dataset that we generated, where each data point $x $
contains $8$ random numbers $\{x_{1},...,x_{8}\}$ uniformly distributed over
the interval $[0,1)$. The intention of using such a random dataset is to
ensure that our experimental results could stand less chances to be biased
by the structure of the dataset.

The other is the $784$-dimensional Modified National Institute of Standards
and Technology (MNIST) dataset \cite{MNIST}, which is a widely-used resource
for machine learning research \cite{MNIST2}. It contains $70000$ greyscale $%
28\times 28=784$ pixel images of handwritten digits $0\symbol{126}9$.

\subsection{The feature map}


Our feature map uses the amplitude encoding method \cite{ml197}, where each $%
d$-dimensional classical input data point $x=\{x_{1},...,x_{d}\}$\ is
encoded as the amplitudes of a $Q$-qubit quantum state%
\begin{equation}
\left\vert \Phi _{x}\right\rangle
=C_{norm}\sum\limits_{i=1}^{d}x_{i}\left\vert i\right\rangle.
\end{equation}%
Here $d=2^{Q}$, $\left\vert i\right\rangle $\ is the $i$th computational
basis state, and $C_{norm}$\ is the normalization constant.


\begin{table*}[t]
\caption{Precision comparison between the two approaches on the obtained
value of the gradient $grad(\protect\theta _{i})$ with respect to the
adjustable parameter $\protect\theta _{i}$\ ($i=1,...,6$). The VQC takes $8$%
-dimensional classical input data, with the \textit{RealAmplitudes} ansatz
containing $r=1$ repetition and full entanglement. The number of input data
is $m=20$. The exact values are calculated numerically. The experimental
values of both approaches are obtained using quantum simulator by taking $%
s=500$. The relative deviation $\Delta$ is calculated as (experimental value
- exact value) / exact value.}
\label{tab:addlabel}\centering
\begin{tabular}{c|c|c|c|c|c}
\hline
\multirow{2}[3]{*}{Gradient} & \multicolumn{1}{c|}{Exact} &
\multicolumn{2}{c|}{Conventional approach} & \multicolumn{2}{c}{Improved
approach} \\ \cline{3-6}
& value & Experimental value & $\Delta$ & Experimental value & $\Delta$ \\
\hline
$grad(\theta _{1})$ & 0.1527 & 0.1589 & 0.0403 & 0.1573 & 0.0298 \\
$grad(\theta _{2})$ & -0.0030 & -0.0050 & 0.6667 & 0.0010 & -1.3181 \\
$grad(\theta _{3})$ & -0.2667 & -0.2691 & 0.0088 & -0.2702 & 0.0131 \\
$grad(\theta _{4})$ & -0.0837 & -0.0897 & 0.0711 & -0.0841 & 0.0052 \\
$grad(\theta _{5})$ & 0.1074 & 0.1064 & -0.0093 & 0.1034 & -0.0369 \\
$grad(\theta _{6})$ & -0.1813 & -0.1755 & -0.0323 & -0.1800 & -0.0073 \\
\hline
\end{tabular}%
\end{table*}


\subsection{The cost function}


Though the form of the cost function plays an important role on the overall
performance of a neural network, it is less relevant to the current work
because our main goal is to test the runtime of the improved approach only,
which depends solely on the structure of the ansatz. Therefore, the cost
function of our networks (either using the improved approach or the
conventional one) is simply taken as%
\begin{equation}
Cost=\frac{1}{m}\sum\limits_{x}\left\vert a-y(x)\right\vert
\end{equation}%
where $m$ is the number of input data, $a$ is the vector of outputs from the
network when $x$ is input, and $y(x)$ is the desired output, i.e., the label
of the input $x$, as contained in the MNIST dataset.

Like the MNIST dataset, we also give each data in our $8$-dimensional input
dataset a label, so that our program can serve as a classifier if needed.
But since our current data are generated randomly, we simply label all data
as \textquotedblleft $2$\textquotedblright\ without actual meaning.

\subsection{Results}


The first experiment is to test the validity of our improved approach. It is
performed on quantum simulator using a VQC with the $8$-dimensional
classical input data, which takes $Q=3$ qubits to encode when using the
amplitude encoding method. The \textit{RealAmplitudes} ansatz in this VQC
contains $1$ repetition with full entanglement. Thus, it has $n=6$
adjustable parameters. We first calculate the exact amplitude distribution
of the output states numerically, and use the result of the gradients as
reference. Then we run the conventional approach for $s=500$ shots. For
comparison, our improved approach for the same VQC is run for $s^{\prime
}=500\times (2n+1)=6500$ shots, so that each cost function could be
calculated for about $s=500$ shots in average (for simplicity, in the
following when we say that the average number of shots for our improved
approach is $s$, we mean that the actual number of shots is $s^{\prime
}=s(2n+1)$). The numbers of input data are all taken to be $m=20$ in these
experiments. The results of the gradients are shown in Table I.

From the results we can see that when the (average) number of shots is $%
s=500 $, both approaches result in similar precision when comparing with the
exact values. (The precision for the gradient $grad(\theta _{2})$ with
respect to the\ second adjustable parameter $\theta _{2}$ is low for both
approaches, probably because the point happens to meet the barren plateau
\cite{10-of-ml277,38-of-ml209,18-of-ml211,ml277}.) Thus, it is proven that
the modified quantum circuit (i.e., Fig.3) in our improved approach works
correctly as desired.

Secondly, we study the individual number of shots for each single cost
function that was actually calculated in the above experiment using our
improved approach. The result is shown in Fig.4. It is found that the
fluctuation of the individual number of shots falls within the range $%
[445,587]$, with a standard deviation $22.6$, which is only $4.52\%$ of the
average value $s=500$, indicating that all the cost functions were
calculated to almost the same precision.


\begin{figure}[t]
\centering
\includegraphics[scale=0.6]{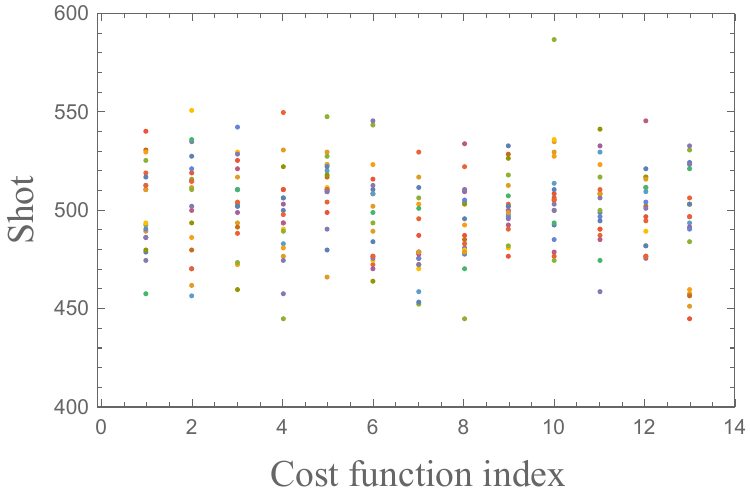}
\caption{Individual number of shots actually computed for each cost function
in our improved approach, corresponding to the experiment shown in the last
column of Table 1. The average number of shots is taken to be $s=500$. The
number of input data is $m=20$. The cost function with index $1$ is the
unshifted cost function, while the other $12$ cost functions are all shifted
ones. The dots with the same color represent the number of shots for the
same input data point.}
\label{fig:epsart}
\end{figure}


Third and most importantly, the runtimes of both approaches are compared
using 3 experiments:

Exp.1: The VQC used in Table I and Fig.4,
run on quantum simulator.

Exp.2: The same VQC 
run on real quantum hardware (the \textquotedblleft
ibmq\_quito\textquotedblright\ backend).

Exp.3: A VQC with $784$-dimensional classical input data, with the \textit{%
RealAmplitudes} ansatz containing $r=2$ repetitions and full entanglement,
run on quantum simulator. Note that this VQC takes $Q=10$ qubits to encode
each input data with the amplitude encoding method, so that the total number
of adjustable parameters is $n=30$ as given by Eq. (\ref{num_param}).


\begin{figure}[tbp]
\centering
\includegraphics[scale=0.6]{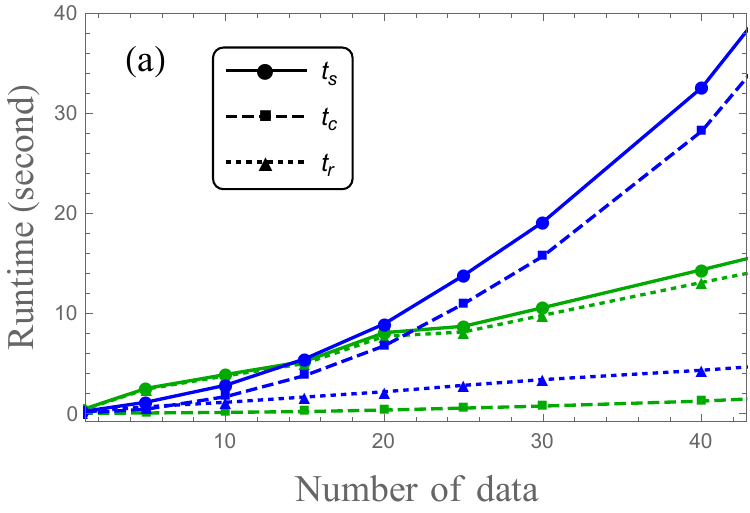} \bigskip
\par
\includegraphics[scale=0.6]{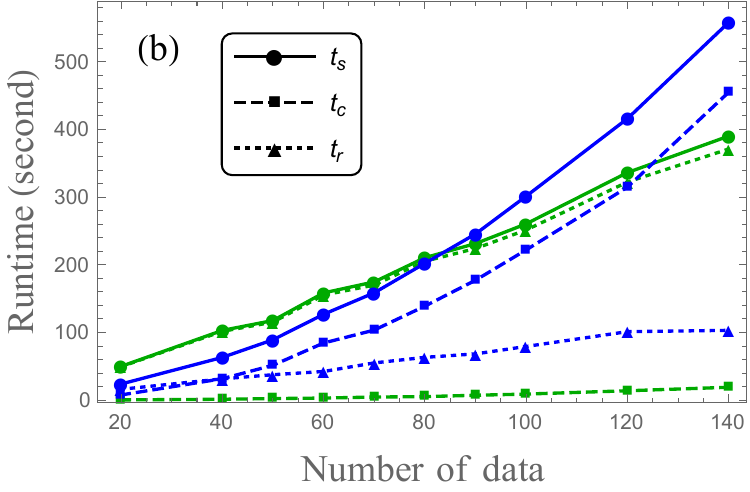} \bigskip
\par
\includegraphics[scale=0.6]{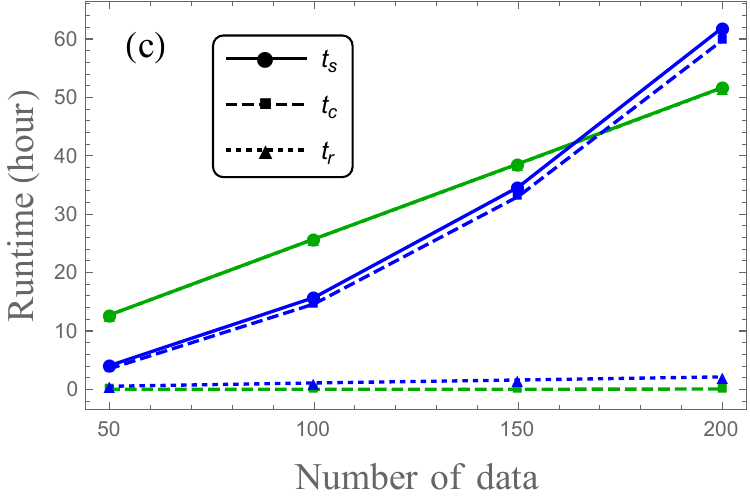}
\caption{Runtimes as a function of the number of input data $m$. The dashed
lines represent the time $t_{c}$ spent on compiling the circuit, the dotted
lines represent the time $t_{r}$ spent on running the circuit, and the solid
lines represent the total runtime $t_{s}=t_{c}+t_{r}$. All blue lines stand
for the conventional approach, and all green lines stand for our improved
approach. (a) Exp.1: The VQC used in the experiment in Fig.4, with $8$%
-dimensional classical input data and the \textit{RealAmplitudes} ansatz
containing $r=1$ repetition and full entanglement, run on quantum simulator.
(b) Exp.2: The same VQC as that of Exp.1, run on real quantum hardware. (c)
Exp.3: A VQC with $784$-dimensional classical input data, with the \textit{%
RealAmplitudes} ansatz containing $r=2$ repetitions and full entanglement,
run on quantum simulator. Note that the green solid and dotted lines almost
overlap with each other since $t_{c}\gg t_{r}$ so that $t_{s}\simeq t_{c}$
for our improved approach in this case.}
\label{fig:epsart}
\end{figure}


Let $t_{c}$ ($t_{r}$) denote the time spent on compiling (running) the
circuit. 
The sum $t_{s}=t_{c}+t_{r}$ can serve as a good measure of the performance
of the approaches, as the other parts of the computer programs (e.g.,
reading the input data and initial parameters, calculating the cost
functions and gradients from the raw counting result of quantum simulator or
real hardware, and exporting the results to the output files) run very fast
and take almost the same amount of time for both approaches. The results are
shown in Fig.5. The number of shots for each data point is fixed as $s=500$
(i.e., the actual number of shots is $s^{\prime }=500(2n+1)$ for our
improved approach).

The following observations can be found in all these experiments.

(1) The values of $t_{r}$\ show that our improved approach always takes more
time to run than the conventional approach does, either on simulator or real
hardware. The difference is about $3\symbol{126}4$ times (in Exp.1 and
Exp.2) to $24$ times (in Exp.3). This is not surprising, because the
improved one needs to be run for more shots to reach the same precision.

(2) On the contrary, the values of $t_{c}$ show that our improved approach
saves a tremendous amount of time for compiling the circuit, which is only
about $4\%$ (in Exp.1 and Exp.2) to $0.1\%$ (in Exp.3) of that of the
conventional approach. This is owed to the significantly reduced circuit
depth, as shown in Eq. (\ref{improved depth}).

(3) As the result of the competition between $t_{c}$ and $t_{r}$, the total
runtime $t_{s}$ of the conventional approach is shorter when $m$ (the number
of input data) is low. But as $m$ increases, our improved approach will
eventually become faster. In Exp.1, the turning point occurs at around $m=15
$ where both approaches take about $8$ seconds to complete. After that, $%
t_{s} $ of our approach increases approximately in a linear way, while $%
t_{s} $ of the conventional one rises dramatically, mostly due to the long
compiling time $t_{c}$. In both Exp.2 and Exp.3, the same behavior can be
observed. The turning points occur relatively later though. This is because
the real quantum hardware used in Exp.2 runs about $10\symbol{126}30$ times
slower than the simulator used in Exp.1, and Exp.3 has a more complicated
circuit than that of Exp.1 though it is also run on simulator. Consequently,
both Exp.2 and Exp.3 have a longer $t_{r}$\ than that of Exp.1. Meanwhile,
all the three experiments have similar $t_{c}$ values because the compiling
of the circuits are always done on classical computers. Therefore, the
slower the quantum hardware or simulator is, the later the turning point of
the total runtime $t_{s}$ occurs.

In the age of big data, for real applications of neural networks, the number
of input data is generally much higher than what was used here. As a result,
we can see that our improved approach has the advantage that it can save the
total runtime. This advantage is expected to be even more obvious with the
advance of real quantum computers, because they will surely run faster in
the future than they do today, resulting in a shorter $t_{r}$ so that the
turning point of $t_{s}$ could\ occur even earlier.

It is also worth noting that if Qiskit's \textquotedblleft
append\textquotedblright\ function were used for stacking the circuits, the
\textquotedblleft ibmq\_quito\textquotedblright\ real quantum backend only
allows $100$ circuits to be appended in a single job. Since the VQC in Exp.2
has $6$ adjustable parameters, there are totally $13$ shifted and unshifted
cost functions to be calculated for each single input data. It means that
using the \textquotedblleft append\textquotedblright\ function can handle
only $m=[100/13]=7$ input data at the most. But Exp.2 (i.e., Fig.5(b)) shows
that both the conventional approach (using the \textquotedblleft
compose\textquotedblright\ function) and our improve approach can accomplish
the computation for at least $m=140$ input data in a single job with no
problem. Thus, they manage to break the limit on real quantum hardware.


\begin{figure}[tb]
\centering
\includegraphics[scale=0.6]{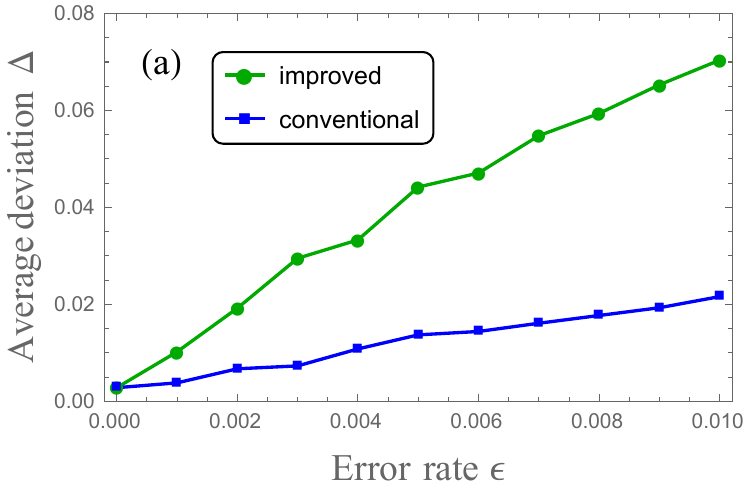} \bigskip
\par
\includegraphics[scale=0.6]{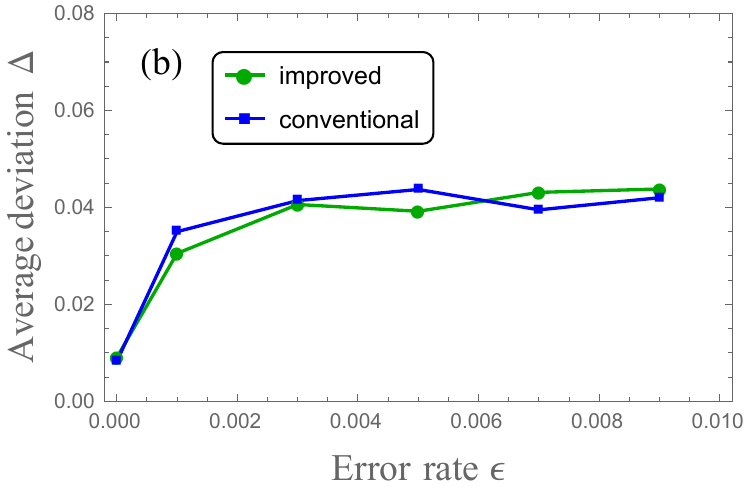}
\caption{The average deviation $\bar{\Delta}$ (as defined in Eq. (\protect
\ref{dev})) of the cost functions as a function of $\protect\epsilon $,
which is the error rate of all quantum gates. The blue lines stand for the
conventional approach, and the green lines stand for our improved approach.
All data are computed using quantum simulator. (a) The same VQC used in
Exp.1 of Fig.5(a), with $8$-dimensional classical input data. The number of
input data for each point in the figure is fixed as $m=20$. (b) A VQC with $%
784$-dimensional classical input data. The number of input data is $m=1$.
Both VQCs use the \textit{RealAmplitudes} ansatz with $r=1$ repetitions and
full entanglement.}
\label{fig:epsart}
\end{figure}


Finally, the impact of noise in the circuits is studied and shown in Fig.6,
where the value of the average deviation $\bar{\Delta}$ is calculated as
follows. The exact values $f_{i}^{exa}$ ($i=1,...,2n+1$) of all the shifted
and unshifted cost functions were calculated numerically beforehand. Then
the values $f_{i}^{sim}$\ of these cost functions are computed using the
noise model in the simulator, where the error rate $\epsilon $ of all
quantum gates are taken to be equal for simplicity, and ranges from 0.001 to 0.01. The number of shots for
each cost function is $s=1000$ (i.e., $s^{\prime }=1000(2n+1)$ for our
improved approach). Define%
\begin{equation}
\Delta _{i}=\left\vert \frac{f_{i}^{sim}-f_{i}^{exa}}{f_{i}^{exa}}\right\vert
\end{equation}%
as the absolute value of the relative deviation between the results of the
simulator and the exact values, and the average deviation is obtained as%
\begin{equation}
\bar{\Delta}=\frac{1}{2n+1}\sum\limits_{i=1}^{2n+1}\Delta _{i}.  \label{dev}
\end{equation}%
We first study the same VQC used in Exp.1 of Fig.5(a), which takes $8$%
-dimensional classical input data, with the \textit{RealAmplitudes} ansatz
containing $r=1$ repetitions and full entanglement. The result is shown in
Fig.6(a), where the number of input data is $m=20$. We can see that the
effect of noise in our improved approach is about $2\symbol{126}3$ times
more significant than that in the conventional approach. Theoretically, this
is not unexpected, because it takes $3$ qubits to encode $8$-dimensional
classical input data in the conventional approach, while our improved
approach adds $2$ more qubits to control the shifts to the parameterized
gates. The same error rate surely brings more errors to $5$ qubits than it
does to $3$ qubits. But we can expect that the difference should become less
significant with higher-dimensional classical input data, which means an
increase on the number of total qubits in the VQC. This is because in our
improved approach, the number of additional qubits is always $2$, regardless
the number of qubits in the original VQC. Indeed, a VQC with $784$%
-dimensional classical input data is studied in Fig.6(b). It is similar to
the VQC in Exp.3 of Fig.5(c), but since the noise model in the simulator
runs much slower than the noiseless one, here we only take $r=1$ repetition
in the \textit{RealAmplitudes} ansatz, so that the total number of
adjustable parameters drops to $n=20$. The number of input data is taken as $%
m=1$. For $784$-dimensional input, our improved approach takes $12$ qubits
in total, while the conventional approach takes $10$ qubits. The result in
Fig.6(b) verifies the conjecture that the noise should display less
difference between the two approaches when the number of qubits becomes
close. In fact, when $\epsilon =0.001$, $0.003$ and $0.005$, the average
deviation of our improved approach is even slightly lower than that of the
conventional approach (which is believed to be the consequence of the
randomness in quantum simulation). The link to the source data of all
figures is provided in the Data Availability section. More details of the
computer programs and the environment is given in 
Appendix B.

\section{Discussion}

In summary, it is demonstrated experimentally that our improved approach
manages to increase the number of gradients that can be computed in each
single job, breaking the limit that can be reached using Qiskit's
\textquotedblleft append\textquotedblright\ function. More importantly, it
has a smaller circuit depth and requires less classical registers for
storing the measurement results when comparing with the conventional
approach using Qiskit's \textquotedblleft compose\textquotedblright\
function. This reduces the time spent for compiling the quantum circuit
significantly. Therefore, the total runtime of the program can also be
saved, especially when the number of input data is high and the structure of
the quantum circuit is complicated.

Though we only demonstrated our improved approach for VQCs where each
parameterized quantum gate $U(\theta )$ satisfies%
\begin{equation}
U(\theta +s)=U(\theta )U(s)  \label{gate}
\end{equation}%
(i.e., the gates covered by the original parameter-shift rule \cite{ml124}),
it can also be modified to compute the gradients for any other quantum
gate that requires the general parameter-shift rules \cite{ml138}. For
example, suppose that the first parameterized gate $RY(\theta _{1})$ on the
qubit $q_{1}$\ in Fig.3 is replaced by a more general gate $U^{\prime
}(\theta _{1})$ that does not satisfy\ Eq. (\ref{gate}). To compute $%
U^{\prime }(\theta +s)$, all we need is to replace the controlled-RY gate $%
CRY(s)$ (controlled by $q_{b}$ and acting on $q_{1}$) in the first block
right behind $RY(\theta _{1})$\ in Fig.3 by two controlled-gates $(U^{\prime
}(\theta _{1}))^{-1}$\ and $U^{\prime }(\theta _{1}+s)$\ in serial, both of
which are also controlled by $q_{b}$ and acting on $q_{1}$. Here $(U^{\prime
}(\theta _{1}))^{-1}$\ is the reverse operation of $U^{\prime }(\theta _{1})$
such that $(U^{\prime }(\theta _{1}))^{-1}U^{\prime }(\theta _{1})=I$. We
can see that once the measurement on $q_{b}$\ before them (i.e., the
measurement operator in the first block of Fig.3) results in $\left\vert
1\right\rangle _{q_{b}}$, then these two gates will be activated. Combining
with the original $U^{\prime }(\theta _{1})$ gate, the complete operation on
the qubit $q_{1}$ will be $U^{\prime }(\theta _{1}+s)(U^{\prime }(\theta
_{1}))^{-1}U^{\prime }(\theta _{1})=U^{\prime }(\theta _{1}+s)$. Thus, the
shift of such a general parameterized quantum gate can also be computed with
our approach.

Our approach also suggests the following improvement on real quantum
computers. From Fig.3 we can see that the purpose of the first two
operations in each block (the controlled-RY gate $CRY(\gamma _{j})$ and the
measurement) is to turn the state of $q_b$ into $\left\vert 1\right\rangle
_{q_{b}}$ with a certain probability. If there is a classical random number
generator (even if it generates pseudorandom numbers only), it can
accomplish the same task while further reducing the circuit depth. Also, the
next two quantum controlled gates make use of the states $\left\vert
0\right\rangle$ and $\left\vert 1\right\rangle$ of $q_a$ and $q_b$ only,
without needing their quantum superpositions. They can be replaced by
classical IF gates (i.e., controlled gates that take classical control bits
instead of control qubits, while the target registers can be either quantum
or classical) and both $q_a$ and $q_b$ can be classical registers too, so
that less quantum resource is required. Unfortunately, classical random
number generator and the IF gate are currently unavailable on real quantum
hardware (the latter is available on the simulator though). Our approach
shows that there is the need for adding these operations in the future, so
that the performance of real quantum computers for certain tasks can be
further improved.

\bigskip 
\noindent \textbf{Acknowledgments}

\noindent This work was supported in part by Guangdong Basic and Applied
Basic Research Foundation under Grant No. 2019A1515011048. The experiments
in this work were supported in part by National Supercomputer Center in
Guangzhou. The author thanks Qing Liu and the technical support team at
National Supercomputer Center.

%

\bigskip \noindent \textbf{Data Availability}

\noindent The raw experimental data of this study are publicly available at
https://github.com/gphehub/grad2210.

\bigskip \noindent \textbf{Code Availability}

\noindent The software codes for generating the experimental data are
publicly available at https://github.com/gphehub/grad2210.



\appendix


\section{Proving that $\protect\lambda $ in Eq. (10) is non-trivial}


In the \textit{RealAmplitudes} ansatz, let $Q$ denote the total number of
the qubits. When $Q$ is fixed, increasing $n$ means increasing the number of
repetition $r$ in this ansatz. This is because the ansatz starts with a
parameterized RY gate for each of the $Q$ qubits, then each repetition adds
another parameterized RY gate to each qubit. Thus, the total number of the
parameters satisfies%
\begin{equation}
n=Q(r+1).  \label{num_param}
\end{equation}%
Suppose that the circuit depth of each additional repetition (including the
RY gates and the CX gates in the entanglement section) is $\Delta $. Then%
\begin{equation}
\Lambda =\Lambda _{0}+r\Delta =\Lambda _{0}+(\frac{n}{Q}-1)\Delta =(\Lambda
_{0}-\Delta )+n\frac{\Delta }{Q},
\end{equation}%
where $\Lambda _{0}$ denotes the depth of the other parts of the circuit
which is not included in $\Delta $, e.g., the feature map, the observable,
the final measurement, and the very first RY gate on each qubit within the
ansatz. In this case, we yield%
\begin{equation}
\lambda =\frac{(\Lambda _{0}-\Delta )+n\frac{\Delta }{Q}}{n}=\frac{(\Lambda
_{0}-\Delta )}{n}+\frac{\Delta }{Q}.
\end{equation}%
A neural network that is useful in practice generally has a large number of
parameters. When $n\rightarrow \infty $\ we have
\begin{equation}
\lim_{n\rightarrow \infty }\lambda =\Delta /Q.
\end{equation}%
Thus it is proven that $\lambda $ stays non-trivial.


\section{Program details}


The programs for $8$-dimensional input data using quantum simulator were run
on a personal laptop with a $2.8$ GHz quad-core Intel i7-3840QM processor
and $16$ GB $1,600$ MHz DDR3 memory.

The programs for $784$-dimensional input data using quantum simulator were
run on a Tianhe-2 supercomputer node with a $2.2$ GHz Intel Xeon E5-2692 v2
processor and $64$ GB memory.

In these simulations, Qiskit's \textquotedblleft
qasm\_simulator\textquotedblright\ backend was used as the quantum simulator.

The programs using real hardware were run on the \textquotedblleft
ibmq\_quito\textquotedblright\ $5$-qubit backend of IBM Quantum Experience
online platform.

The operating environment for all programs is Python 3.8.10 with Qiskit
0.21.2.

When running on quantum simulator, both $t_{c}$ (the time spent on compiling
the quantum circuit) and $t_{r}$ (the time spent on running the circuit) are
recorded automatically by the programs. But when running on real quantum
hardware, $t_{r}$ is obtained from the job output files downloaded from IBM
Quantum Experience platform instead. This is because the value of $t_{r}$
recorded by the programs in this case also includes the time spent on
waiting in the queue for the job to be run, which is not the correct measure
for the performance of the approaches being used.

In both approaches, the circuits for different input data are stacked
together using Qiskit's \textquotedblleft compose\textquotedblright\
function.

The initial values of all adjustable parameters of the VQCs were generated
randomly (uniformly distributed over the interval $[0,\pi )$) and stored
beforehand. Then we always loaded these same values into every computer
programs in every run.

More details can be found in the file \textquotedblleft
file\_description.pdf\textquotedblright\ accompanies with the source codes
(see Code Availability section).

%
%

%


\begin{thebibliography}{99}
\bibitem{ml179} Cerezo, M., Verdon, G., Huang, H. -Y., Cincio, L. \& Coles,
P. J. Challenges and opportunities in quantum machine learning. \textit{Nat.
Comput. Sci.} \textbf{2}, 567--576 (2022).

\bibitem{back} Rumelhart, D. E., Hinton, G. E. \& Williams, R. J. Learning
representations by back-propagating errors. \textit{Nature} \textbf{323},
533--536 (1986).

\bibitem{ml98} Nielson, M. A. \textit{Neural networks and deep learning}
(Determination Press, 2015).

\bibitem{ml211} Abbas, A., King, R., Huang, H. -Y., Huggins, W. J.,
Movassagh, R., Gilboa, D. \& McClean, J. R. On quantum backpropagation,
information reuse, and cheating measurement collapse. Preprint at
https://arxiv.org/abs/2305.13362 (2023).

\bibitem{ml209} Bowles, J., Wierichs, D. \& Park, C. -Y. Backpropagation
scaling in parameterised quantum circuits. Preprint at
https://arxiv.org/abs/2306.14962 (2023).

\bibitem{ml124} Schuld, M., Bergholm, V., Gogolin, C., Izaac, J. \&
Killoran, N. Evaluating analytic gradients on quantum hardware. \textit{%
Phys. Rev. A} \textbf{99}, 032331 (2019).

\bibitem{ml138} Wierichs, D., Izaac, J., Wang, C. \& Lin, C. Y. -Y. General
parameter-shift rules for quantum gradients. \textit{Quantum} \textbf{6},
677 (2022).

\bibitem{amazon} Amazon braket, https://aws.amazon.com/
cn/braket/pricing
(Access on Jan. 30, 2025).

\bibitem{example} Games \& Demos: Local reality and the CHSH inequality, in
\textit{Qiskit Textbook}.
https://learn.qiskit.org/course/ch-demos/local-reality-and-the-chsh-inequality (Access on Jan. 30, 2023).

\bibitem{IBM} IBM cloud's pay-as-you-go plan, https://www.ibm.com/quantum/access-plans (Access on Jan. 30, 2025).

\bibitem{ml128} Peruzzo, A., McClean, J., Shadbolt, P., Yung, M. -H., Zhou,
X. -Q., Love, P. J., Aspuru-Guzik, A. \& O'Brien, J. L. A variational
eigenvalue solver on a photonic quantum processor. \textit{Nat. Commun.}
\textbf{5}, 4213 (2014).

\bibitem{ml383} Grant, E., Benedetti, M., Cao, S., Hallam, A., Lockhart, J.,
Stojevic, V., Green, A. G. \& Severini, S. Hierarchical quantum classifiers.
\textit{npj Quantum Inf.} \textbf{4}, 65 (2018).

\bibitem{ml54} Benedetti, M., Lloyd, E., Sack, S. \& Fiorentini, M.
Parameterized quantum circuits as machine learning models. \textit{Quantum
Sci. Technol.} \textbf{4}(4), 043001 (2019).

\bibitem{ml118} LaRose, R. \& Coyle, B. Robust data encodings for quantum
classifiers. \textit{Phys. Rev. A} \textbf{102}(3), 032420 (2020).

\bibitem{ml52} Cerezo, M., Arrasmith, A., Babbush, R., Benjamin, S. C.,
Endo, S., Fujii, K., McClean, J. R., Mitarai, K., Yuan, X., Cincio, L. \&
Coles, P. J. Variational quantum algorithms. \textit{Nat. Rev. Phys.}
\textbf{3}, 625--644 (2021).

\bibitem{ml53} Abbas, A., Sutter, D., Zoufal, C., Lucchi, A., Figalli, A. \&
Woerner, S. The power of quantum neural networks. \textit{Nat. Comput. Sci.}
\textbf{1}, 403--409 (2021).

\bibitem{ml32} Arthur, D. \& Date, P. A hybrid quantum-classical neural
network architecture for binary classification. Preprint at
https://arxiv.org/abs/2201.01820 (2022).

\bibitem{ml208} Du, Y., Huang, T., You, S., Hsieh, M. -H. \& Tao, D. Quantum
circuit architecture search for variational quantum algorithms. \textit{npj
Quantum Inf.} \textbf{8}, 62 (2022).

\bibitem{ml368} O'Malley, P.\thinspace J.\thinspace J., Babbush, R.,
Kivlichan, I. D., Romero, J., McClean, J. R., Barends. R., Kelly, J.,
Roushan, P., et al. Scalable quantum simulation of molecular energies.
\textit{Phys. Rev. X} \textbf{6}(3), 031007 (2016).

\bibitem{ml157} Kandala, A., Mezzacapo, A., Temme, K., Takita, M., Brink,
M., Chow, J. M. \& Gambetta, J. M. Hardware-efficient variational quantum
eigensolver for small molecules and quantum magnets. \textit{Nature} \textbf{%
549}, 242--246 (2017).

\bibitem{ml367} C\^{\i}rstoiu, C., Holmes, Z., Iosue, J., Cincio, L., Coles,
P. J. \& Sornborger, A. Variational fast forwarding for quantum simulation
beyond the coherence time. \textit{npj Quantum Inf.} \textbf{6}, 82 (2020).

\bibitem{ml366} Kim, Y., Eddins, A., Anand, S., Wei, K. X., van den Berg,
E., Rosenblatt, S., Nayfeh, H., Wu, Y., Zaletel, M., Temme, K. \& Kandala,
A. Evidence for the utility of quantum computing before fault tolerance.
\textit{Nature} \textbf{618}, 500--505 (2023).

\bibitem{ml382} Li, Y. \& Benjamin, S. C. Efficient variational quantum
simulator incorporating active error minimization. \textit{Phys. Rev. X}
\textbf{7}(2), 021050 (2017).

\bibitem{ml381} Endo, S., Sun, J., Li, Y., Benjamin, S. C. \& Yuan, X.
Variational quantum simulation of general processes. \textit{Phys. Rev. Lett.%
} \textbf{125}(1), 010501 (2020).

\bibitem{ml30} Farhi, E., Goldstone, J. \& Gutmann, S. A quantum approximate
optimization algorithm. Preprint at https:// arxiv.org/abs/1411.4028 (2014).

\bibitem{ml380} Zhou, L., Wang, S. -T., Choi, S., Pichler, H. \& Lukin, M.
D. Quantum approximate optimization algorithm: Performance, mechanism, and
implementation on near-term devices. \textit{Phys. Rev. X} \textbf{10}(2),
021067 (2020).

\bibitem{ml387} Farhi, E., Goldstone, J., Gutmann, S. \& Zhou, L. The
quantum approximate optimization algorithm and the sherrington-kirkpatrick
model at infinite size. \textit{Quantum} \textbf{6}, 759 (2022).

\bibitem{161-of-ml224} Terashi, K., Kaneda, M., Kishimoto, T., Saito, M.,
Sawada, R. \& Tanaka, J. Event classification with quantum machine learning
in high-energy physics. \textit{Comput. Softw. Big Sci.} \textbf{5}, 2
(2021).

\bibitem{ml379} Gianelle, A., Koppenburg, P., Lucchesi, D., Nicotra, D.,
Rodrigues, E., Sestini, L., de Vries, J. \& Zuliani, D. Quantum machine
learning for $b$-jet charge identification. \textit{J. High Energ. Phys.}
\textbf{2022}, 14 (2022).

\bibitem{159-of-ml224} Islam, M., Chowdhury, M., Khan, Z. \& Khan, S. M.
Hybrid quantum-classical neural network for cloud-supported in-vehicle
cyberattack detection. \textit{IEEE Sens. Lett.} \textbf{6}(4), 6001204
(2022).

\bibitem{ml389} Yusuf K\"{u}\c{c}\"{u}kkara, M., Atban, F. \& Bay\i lm\i
\c{s} C. Quantum-neural network model for platform independent DDos attack
classification in cyber security. \textit{Adv. Quantum Technol.} \textbf{7}%
(10), 2400084 (2024).

\bibitem{160-of-ml224} Emmanoulopoulos, D. \& Dimoska, S. Quantum machine
learning in finance: Time series forecasting. Preprint at
https://arxiv.org/abs/2202.00599 (2022).

\bibitem{ml386} Cherrat, E. A., Raj, S., Kerenidis, I., Shekhar, A., Wood,
B., Dee, J., Chakrabarti, S., Chen, R., et al. Quantum deep hedging. \textit{%
Quantum} \textbf{7}, 1191 (2023).

\bibitem{ansatz} RealAmplitudes, in \textit{Qiskit Documentation}. https://qiskit.org/documentation/
stubs/qiskit.circuit.library.RealAmplitudes.
html
(Access on Oct. 6, 2022).

\bibitem{MNIST} LeCun, Y. The MNIST database of handwritten digits.
http://
yann.lecun.com/exdb/mnist/ (1998).

\bibitem{MNIST2} Deng, L. The MNIST database of handwritten digit images for
machine learning research. \textit{IEEE Signal Processing Magazine} \textbf{%
29}, 141--142 (2012).

\bibitem{ml197} LaRose, R. \& Coyle, B. Robust data encodings for quantum
classifiers. \textit{Phys. Rev. A} \textbf{102}, 032420 (2020).

\bibitem{10-of-ml277} McClean, J. R., Boixo, S., Smelyanskiy, V. N.,
Babbush, R. \& Neven, H. Barren plateaus in quantum neural network training
landscapes. \textit{Nat. Commun.} \textbf{9}, 4812 (2018).

\bibitem{38-of-ml209} Grant, E., Wossnig, L., Ostaszewski, M. \& Benedetti,
M. An initialization strategy for addressing barren plateaus in parametrized
quantum circuits. \textit{Quantum} \textbf{3}, 214 (2019).

\bibitem{18-of-ml211} Wang, S., Fontana, E., Cerezo, M., Sharma, K., Sone,
A., Cincio, L. \& Coles, P. J. Noise-induced barren plateaus in variational
quantum algorithms. \textit{Nat. Commun.} \textbf{12}, 6961 (2021).

\bibitem{ml277} Sack, S. H., Medina, R. A., Michailidis, A. A., Kueng, R. \&
Serbyn, M. Avoiding barren plateaus using classical shadows. \textit{PRX
Quantum} \textbf{3}(2), 020365 (2022).
\end{thebibliography}
\end{document}